\documentclass[aps,prb,reprint,amssymb,amsmath,superscriptaddress]{revtex4-1}

\usepackage{graphicx}

\begin{document}

\title{Two-dimensional semimetal in a wide HgTe quantum well:  magnetotransport and energy spectrum }

\author{G.~M.~Minkov}
\affiliation{Institute of Metal Physics RAS, 620990 Ekaterinburg,
Russia}

\affiliation{Institute of Natural Sciences, Ural Federal University,
620000 Ekaterinburg, Russia}

\author{A.~V.~Germanenko}

\author{O.~E.~Rut}
\affiliation{Institute of Natural Sciences, Ural Federal University,
620000 Ekaterinburg, Russia}

\author{A.~A.~Sherstobitov}
\affiliation{Institute of Metal Physics RAS, 620990 Ekaterinburg,
Russia}

\affiliation{Institute of Natural Sciences, Ural Federal University,
620000 Ekaterinburg, Russia}

\author{S.~A.~Dvoretski}

\affiliation{Institute of Semiconductor Physics RAS, 630090
Novosibirsk, Russia}

\author{N.~N.~Mikhailov}

\affiliation{Institute of Semiconductor Physics RAS, 630090
Novosibirsk, Russia}

\date{\today}

\begin{abstract}
The results of experimental study of the magnetoresistivity, the Hall
and Shubnikov-de Haas effects for the heterostructure with  HgTe
quantum well of $20.2$~nm width are reported. The measurements were
performed on the gated samples over the wide range of electron and hole
densities including vicinity of a charge neutrality point. Analyzing
the data we conclude that the energy spectrum is drastically different
from that calculated in framework of $kP$-model. So, the hole effective
mass is equal to approximately $0.2\,m_0$ and practically independent
of the quasimomentum ($k$) up to $k^2\gtrsim 0.7\times
10^{12}$~cm$^{-2}$, while the theory predicts negative (electron-like)
effective mass up to $k^2=6\times 10^{12}$~cm$^{-2}$.  The experimental
effective mass near $k=0$, where the hole energy spectrum is
electron-like, is close to $-0.005 m_0$, whereas the theoretical value
is about $-0.1\, m_0$.
\end{abstract} \pacs{73.20.Fz, 73.61.Ey}

\maketitle

\section{Introduction}
\label{sec:intr}

Two-dimensional (2D) systems based on gapless semiconductors such as
HgTe represent unique object. A great variety of two-dimensional
electron and hole systems based on this materials can be realized
depending on the quantum well width ($d$) and content of cadmium in the
well and barriers Hg$_{1-x}$Cd$_x$Te. It is well established now that
the energy spectrum in single CdTe/HgTe/CdTe quantum well  at
$d=d_c\simeq 6.5$~nm is gapless\cite{Gerchikov90} and close to the
linear Dirac-like spectrum  at small quasimomentum
($k$).\cite{Bernevig06} When thickness $d<d_c$ (i.e., when the HgTe
quantum well is narrow) the energy spectrum is analogous to the
spatially quantized spectrum of narrow gap semiconductor such as InSb.
For thick HgTe layer, $d>d_c$, the quantum well is in inverted regime
when the main electron subband of spatial quantization is formed at
$k=0$ from the heavy hole states.\cite{Dyak82e}

The energy spectrum and transport phenomena of 2D carriers in HgTe
based structures were studied intensively last decade both
experimentally\cite{Landwehr00,Zhang02,Ortner02,Zhang04,Koenig07,Gusev11,Kvon11}
and theoretically.\cite{Bernevig06,Tkachov11,Nicklas11,Ostrovsky12} The
experimental data on the energy distance between the different 2D
subbands at zero quasimomentum are in satisfactory agreement with the
theory. Electron energy spectrum, electron effective mass and their
dependence on the quantum well width are in agreement with the
calculations also. As regards to the experimental data on the valence
band energy spectrum, namely the value of bands overlapping, role of
strain, effective masses at $k=0$ and at large quasimomentum, they are
discrepant and call for further investigation.

In this paper, we present the results of experimental study of the
transport properties of the heterostructure with the HgTe quantum well
with the inverted energy spectrum. The measurements were performed over
wide range of electron and hole densities including the  vicinity of
the charge neutrality point (CNP) with nearly equal electron and hole
densities. Analysis of experimental data brings us to the picture of
the energy spectrum, which drastically differs from the commonly
accepted one.

\section{Experimental}
\label{sec:expdet}

\begin{figure}
\includegraphics[width=\linewidth,clip=true]{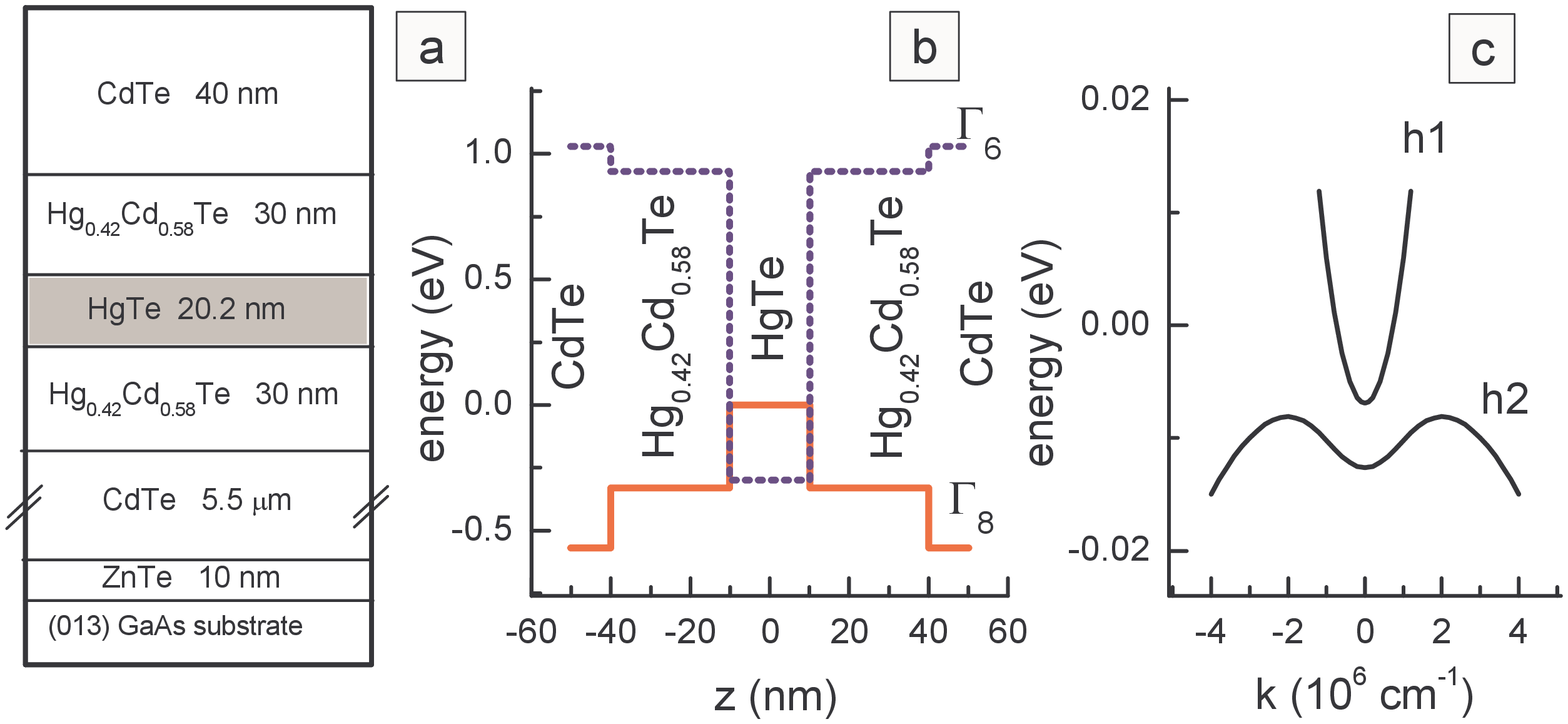}
\caption{(Color online) Architecture (a) and energy  diagram (b)
of the structure under investigation. (c) -- The energy spectrum
calculated within the framework of the isotropic six-band $kP$-model.}\label{f1}
\end{figure}

Our HgTe quantum wells  were realized on the basis of
HgTe/Hg$_{1-x}$Cd$_{x}$Te ($x=0.58$) heterostructure grown by means of
MBE on GaAs substrate with the (013) surface
orientation.\cite{Mikhailov06} The nominal width of the quantum well
was   $d=20.2$~nm. The samples were mesa etched into standard Hall
bars. To change and control the electron and hole densities ($n$ and
$p$, respectively) in the quantum well, the field-effect transistors
were fabricated with the parylene as an insulator and aluminium as a
gate electrode. The measurements were performed at the temperature of
liquid helium in the magnetic field up to $8$~T. All the data will be
presented for $T=1.35$~K, unless otherwise specified. The architecture
and the energy diagram of the structure investigated is shown in
Fig.~\ref{f1}(a) and Fig.~\ref{f1}(b), respectively. The sketch of
energy spectrum calculated within framework of the isotropic six-band
$kP$-model with taking into account the lattice mismatch is presented
in Fig.~\ref{f1}(c). One can see that h1-to-h2 interband distance at
$k=0$ is about $5$~meV, the dispersion $E(k)$  in the valence band is
non-monotonic.

\section{Results and discussion}
\label{sec:res}

\begin{figure}
\includegraphics[width=\linewidth,clip=true]{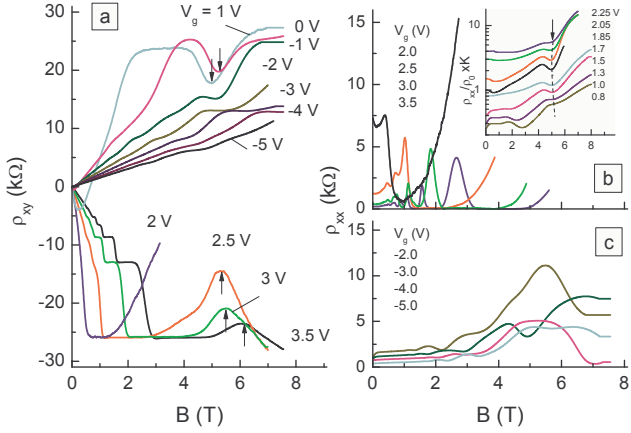}
\caption{(Color online) The magnetic field dependences of  $\rho_{xy}$ (a) and  $\rho_{xx}$ (b, c)
measured for the different gate voltages.  The minimum resulting from the
crossing between the (h1, $n=0$) and (h2, $n=2$) Landau levels is marked by arrow
(for more details, see Section~\ref{sec:cnp}).
The inset in (b) illustrates the weak sensitivity
of the minimum position to the gate voltage near the charge neutrality point,
$V_g\simeq 1.8$~V.}\label{f2}
\end{figure}

An overview of the magnetic field dependences of a transverse
($\rho_{xy}$) and longitudinal ($\rho_{xx}$)  resistivity for different
gate voltages ($V_g$) is presented in Fig.~\ref{f2}. It is seen that
well defined quantum Hall plateaus of $\rho_{xy}$ and minimum of
$\rho_{xx}$ are observed at electron ($V_g>2$~V) and hole ($V_g<-1$~V)
conductivity. Two peculiarities of these dependences should be pointed
out. First, some minimum on the $\rho_{xy}$~versus~$B$ and
$\rho_{xx}$~versus~$B$ dependences at $B=(4-6)$~T [marked by arrows in
Fig.~\ref{f2}(a) and in the inset in Fig.~\ref{f2}(b)] is observed. Its
position only slightly depends on $V_g$  within the gate voltage range
from $3.0$~V to $0.8$~V.

\begin{figure}
\includegraphics[width=0.8\linewidth,clip=true]{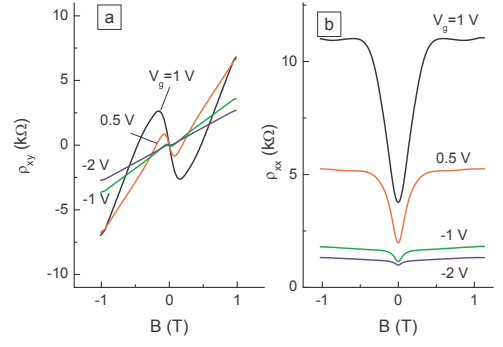}
\caption{(Color online) The low-magnetic-field dependences of $\rho_{xy}$ (a) and $\rho_{xx}$ (b)
measured for $B<1$~T at different gate voltages near the charge neutrality point.}\label{f3}
\end{figure}
\begin{figure}
\includegraphics[width=0.8\linewidth,clip=true]{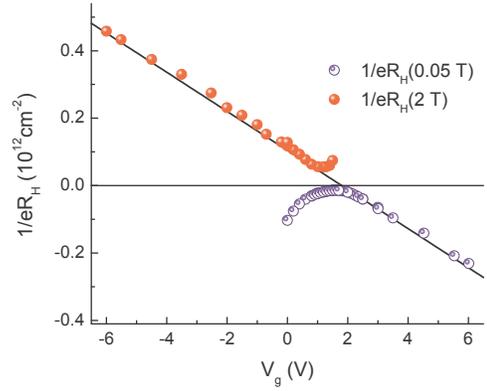}
\caption{(Color online) The gate voltage dependence of $1/eR_H(0.05\text{~T})$
and $1/eR_H(2\text{~T})$. The line is the charge density in the quantum well
calculated as $C(1.8\text{~V}-V_g)/e$, where $C$ is the  capacity between the gate electrode
and quantum well measured experimentally, $C=9.1$~nF/cm$^2$.}\label{f4}
\end{figure}

Second, the alternative sign Hall resistivity $\rho_{xy}$, negative at
low magnetic fields $B<(0.1-0.5)$~T and positive at higher ones [see
Fig.~\ref{f3}(a)], is observed within the gate voltage range $(-3\ldots
+1.8)$~V. Such behavior of $\rho_{xy}$ accompanied by strong positive
magnetoresistivity [Fig.~\ref{f3}(b)] shows that two types of carriers,
electrons and holes, take part in the conductivity within this gate
voltage range. The Hall densities of electrons and holes found as
$1/eR_H(B)$ at $B=0.05\text{ T}$ and $B=2\text{ T}$, respectively, are
plotted against the gate voltage in Fig.~\ref{f4}. Excepting the gate
voltage range from $0$~V to $2$~V, the data points fall on a straight
line with the  slope $-5.5\times 10^{10}$~cm$^{-2}$V$^{-1}$, which is
close to $-1/eC$, where $C=9.1$~nF/cm$^2$ is the capacity between the
2D gas and gate electrode measured on the same structure.\footnote{The
capacity measured is practically independent of the gate voltage over
the entire $V_g$ range. Only one percent variation of $C$ resulting
from the finite value of compressibility of the electron gas is evident
at $V_g\approx 1.8$~V.} So, beyond the range $V_g\simeq (0\ldots2)$~V,
$-1/eR_H(0.05\text{ T})$ and $1/eR_H(2\text{ T})$ give the electron and
hole densities, $n$ and $p$, respectively. The gate voltage
$V_g=1.8$~V, at which the straight line crosses zero, corresponds to
CNP.

It is clear that both the peculiarities may result from specific
features of the energy spectrum of 2D carries in the structures under
study. There are several
papers\cite{Landwehr00,Zhang01,Zhang02,Ortner02, Kvon11,Orlita11} on
the energy spectrum of electrons and holes in HgTe quantum wells with
approximately the same width of the well. However, the energy spectrum,
especially of the valence band, is not understood up to now. Therefore,
let us start from analysis of the Shubnikov-de Haas (SdH) oscillations.

\begin{figure}
\includegraphics[width=\linewidth,clip=true]{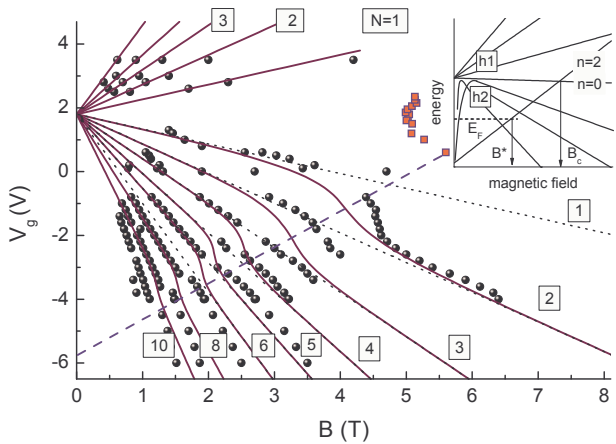}
\caption{(Color online) The fan-chart diagram showing the positions
of the minima in $\rho_{xx}$~versus~$B$ dependence. Symbols
are the experimental results, the squares correspond to the minima
labeled in Fig.~\ref{f2} by arrows. The dashed line is the $B$ dependence of the gate voltage corresponding
to a crossing of the Landau level (h2, $n=2$) with the Fermi level. Solid lines show the expected minima
positions found as $V_g^N=\pm eBN/5.5\times 10^{10}h+1.8$~V (above dashed line)
and $V_g^N=-eB(N+1)/5.5\times 10^{10}h+1.8$~V (below dashed line). The inset is schematic
dispersion of Landau levels.
}\label{f5}
\end{figure}

The positions of the oscillation minima are plotted in the $(B,V_g)$
plane in Fig.~\ref{f5}. This figure resembles a fan-chart showing the
energies of Landau levels as a function of magnetic field. However, it
should be noted that the gate voltage,  rather than the energy is
plotted in the vertical axis. Only in the case when the carrier
effective mass does not depend on the energy, the energy  varies in
direct proportion with $V_g$, $E\propto C V_g/e\nu$, where $\nu$ is
density of states. It is clearly seen that the points laying above the
dashed line fall on the straight lines, which are extrapolated to
$V_g=(1.8\pm 0.1)$~V when $B\to0$ corresponding to CNP. Just such the
behavior should be observed when the variation of density of carriers
with $V_g$ is determined by the geometrical capacity only.\cite{Note1}
The behavior of $\rho_{xx}$ minima near the dashed line will be
discussed below.

\begin{figure}
\includegraphics[width=\linewidth,clip=true]{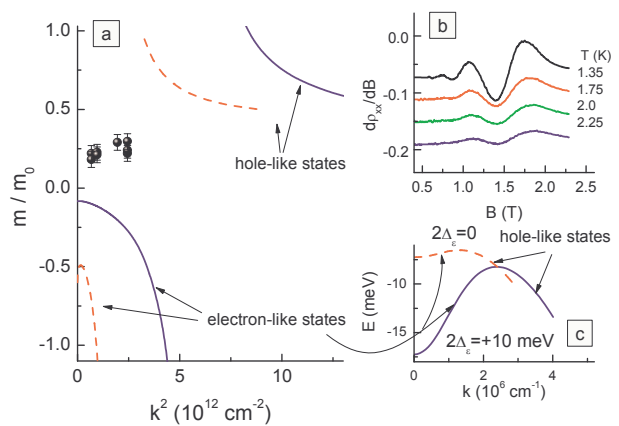}
\caption{(Color online) (a) -- The hole effective mass plotted against the
$k^2$ value as obtained experimentally (symbols) and calculated theoretically
from six-band $kP$-model with and without taking into account the lattice mismatch (the sold and
dashed curves, respectively). (b) -- An example of the SdH oscillations measured for $V_g=0.25$~V
($p=1.05\times 10^{11}$~cm$^{-2}$) at different temperatures.  (c) -- The dispersion for the upper hole subband h2
calculated from six-band $kP$-model with and without taking into account
the lattice mismatch (the solid and dashed curves, respectively).  }\label{f6}
\end{figure}

At hole density higher  than $10^{11}$~cm$^{-2}$ that corresponds to
$V_g< 0$~V, one can find the range of low magnetic field, where the
spin-unsplit SdH oscillations are observed [for example see
Fig.~\ref{f6}(b)]. Fitting the temperature dependence of oscillation
amplitude to the Lifshitz-Kosevich formula,\cite{LifKos55} we have
found the hole effective mass. We succeeded in such analysis within the
density range $(1\ldots4)\times10^{11}$~cm$^{-2}$. The results are
plotted in Fig.~\ref{f6}(a).  One can see that the hole effective mass
is equal to $m_h=(0.2\pm 0.05)\,m_0$ and only slightly increases with
the increasing hole density. Note that the hole density found from SdH
oscillations is close to that found as $1/eR_H(2\text{ T})$.

Let us compare  this result with the result of the conventional
calculation performed within framework of isotropic envelop function
approximation based on six-band $kP$-Hamiltonian. Because the valence
band spectrum noticeably depends on the strain, we present in
Fig.~\ref{f6} the $E$~versus~$k$ and $m_h$~versus~$k$ dependences
calculated for two cases: with and without taking into account the
strain caused by the HgTe and Hg$_{1-x}$Cd$_x$Te lattice mismatch. The
strain effect is characterized by the quantity $2\Delta_\epsilon$,
which is the splitting of $\Gamma_8$ band at $k=0$ in the bulk HgTe.
The estimate for our case, $x=0.58$, gives $2\Delta_\epsilon\simeq
10$~meV. It is seen that the theory for both cases predicts so called
``Mexican hat'' hole energy spectrum, characterized by the
electron-like dispersion $E(k)$ with the positive curvature near $k=0$.
It is significant that the hole effective mass calculated theoretically
is negative up to $k^2\simeq 2\times 10^{12}$~cm$^{-2}$ or
$\simeq6\times 10^{12}$~cm$^{-2}$ depending on the strain.
Experimentally, $m_h$ is positive when $k^2\gtrsim0.7\times
10^{12}$~cm$^{-2}$. This conclusion follows immediately from the fact
that the oscillation minima at $V_g<0$ shift to the more negative $V_g$
with the growing magnetic field (Fig.~\ref{f5}). Note that found values
of $m_h$ are close to those found from cyclotron
resonance.\cite{Kvon11-1} Thus, experimentally found hole effective
mass at low $k$ differs drastically from the calculated one to the
extend that they are different in sign.

The electron effective mass $m_e$ measured by the same way for
$n=(0.6\ldots 1.5)\times 10^{11}$~cm$^{-2}$ is equal to $(0.02\pm
0.005)\,m_0$ that also coincides with the result obtained in the
cyclotron resonance experiments.\cite{Kvon11-1} Such the value of $m_e$
is something less than the calculated one, which is equal to
$0.028\,m_0$ and practically independent of the density up to
$n=3\times 10^{11}$~cm$^{-2}$.

The results discussed above do not give information on the energy
spectrum at small  $k$ values, $k^2\lesssim 0.7\times
10^{12}$~cm$^{-2}$, and on overlapping value of the conduction and
valence bands. Such information can be obtained from analysis of the
low-field magnetoresistivity and Hall effect near CNP.

\begin{figure}
\includegraphics[width=\linewidth,clip=true]{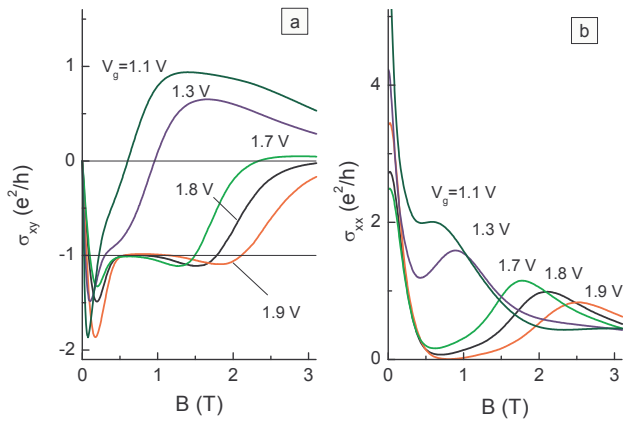}
\caption{(Color online) The magnetic field dependences of $\sigma_{xy}$ (a) and $\sigma_{xx}$ (b) for
several gate voltages near CNP.}\label{f7}
\end{figure}

\section{Transport near CNP. Two types of charge carriers}
\label{sec:cnp}

The detailed measurements at low magnetic field  show, that over the
gate voltage range $(-6 \ldots 1.7)$~V  when the electron density is
less than $(1.0\ldots 1.5)\times 10^{10}$~cm$^{-2}$ and the hole
density is less than $5\times 10^{11}$~cm$^{-2}$, the Hall resistivity
$\rho_{xy}$ is strongly nonlinear in the magnetic field at
$B<(0.1\ldots0.5)$~T insofar that it changes the sign [see
Fig.~\ref{f3}(a)]. The large positive magnetoresistivity is observed
within this magnetic field range [Fig.~\ref{f3}(b)]. Both these facts
strongly suggest that two types of carriers, electrons and holes, take
part in transport, i.e., the system is in two type carrier conductivity
(TTCC) regime. It is more instructive for this case to plot the
magnetic field dependence of $\sigma_{xy}$ and $\sigma_{xx}$ instead of
$\rho_{xx}$ and $\rho_{xy}$ because just the conductivity tensor
components are additive. These dependences for the different gate
voltages are plotted in Fig.~\ref{f7}. It is worth noting that at
$V_g=1.7$~V $\sigma_{xy}$ changes the sign from electron to hole one at
$B=2.2$~T therewith the last electron plateau of $\sigma_{xy}$ (with
$\sigma_{xy}=e^2/h$) is observed at lower magnetic field, $B=
(0.5-1.5)$~T. Such unusual quantum Hall effect determined by minority
carriers, which are electrons in this situation, is observed down to
$V_g=1.4$~V.

Let us analyze the results relating to  the gate voltage range where
two types of carriers take part in the transport in more detail. There
are several physical reasons for such regime: (i) the existence of the
edge states,\cite{Bernevig06,Koenig07,Roth09} which give the electron
contribution to the conductivity, while the 2D gas is of hole type;
(ii) existence of the electron and hole drops due to potential
fluctuations; (iii) the overlapping between the conduction and valence
bands following from the standard $kP$-model.\cite{Kvon11}

\begin{figure}
\includegraphics[width=0.78\linewidth,clip=true]{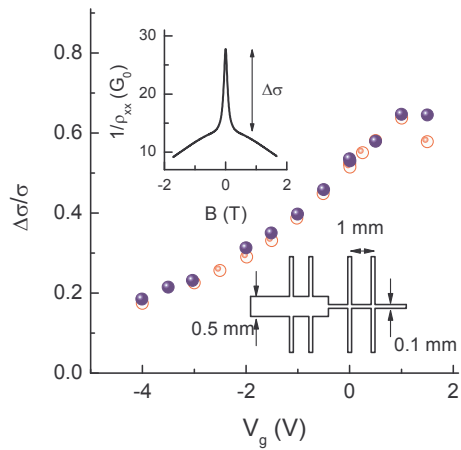}
\caption{(Color online) The relative value of the drop in the
$1/\rho_{xx}$ dependence plotted as a function of the gate voltage for
the wide (open symbols) and narrow (solid symbols) parts of the samples
shown in the sketch. The inset is $1/\rho_{xx}$
plotted as a function of  the magnetic field. Arrow indicates the
drop $\Delta\sigma$.}\label{f8}
\end{figure}

If the edge states result in the conductivity by two types of carriers,
their relative contribution can be estimated from the value of
step-like  drop $\Delta\sigma=1/\rho_{xx}(0.5\text{ T})-1/\rho_{xx}(0)$
evident in the $1/\rho_{xx}$~versus~$B$ dependence (see inset in
Fig.~\ref{f8}). Because this drop should be inversely proportional to
the channel width for this mechanism, we have measured
magnetoresistance of wide ($0.5$~mm) and narrow ($0.1$~mm) parts of the
Hall bar sketched in Fig.~\ref{f8}. As seen from Fig.~\ref{f8} the
relative drop values, $\Delta\sigma/\sigma$, are practically identical
for the wide and narrow parts of the bar. It means that the edge states
in the structures investigated do not responsible for the two types
carriers conductivity regime.

Another possible reason of two type charge carrier conductivity is that
the large long-range fluctuation potential leads to existence of $n$
(or $p$) type droplets in the $p$ (or $n$) type matrix. Due to large
transparency  of $p$-$n$ junction in the structure with inverted energy
spectrum, such medium will demonstrate  features typical for systems
with two types of carriers. The upper limit of the fluctuation
potential amplitude can be estimated from the value of the Dingle
temperature ($T_D$). As follows from analysis of the SdH oscillations
its value does not exceed $1$~meV. So, this mechanism may account for
two type carrier conductivity within narrow gate voltage range only,
which can be estimated as $\Delta V_g\simeq \nu_h T_D
(dn/dV_g)^{-1}\simeq 1$~V ($\nu_h=m_h/\pi\hbar^2$ is the hole density
of states, $m_h=0.2m_0$). Experimentally, this regime occurs within
much wider range of $V_g$: from $-6$~V to $+1.7$~V.

\begin{figure}
\includegraphics[width=\linewidth,clip=true]{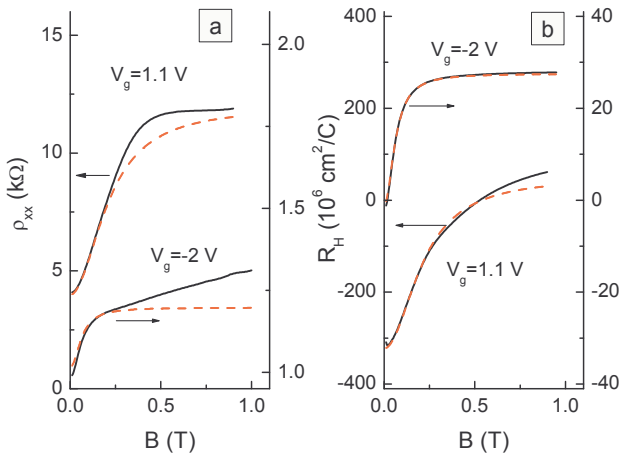}
\caption{(Color online) The magnetic field dependences of $\rho_{xx}$ (a) è $R_H$ (b).
The solid curves are measured experimentally, the dashed lines are the results
of the best fit to the classical formula for TTCC regime. }\label{f9}
\end{figure}

\begin{figure}
\includegraphics[width=\linewidth,clip=true]{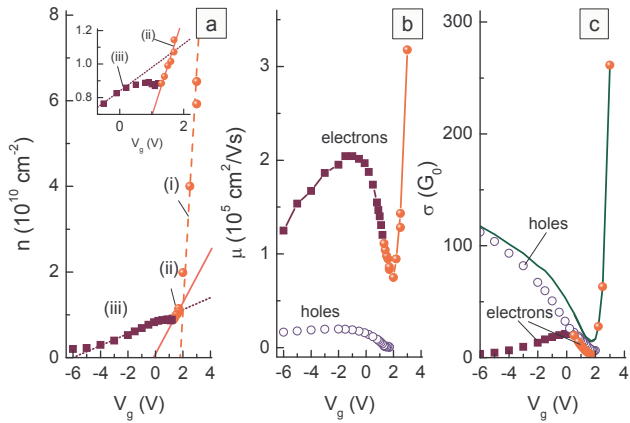}
\caption{(Color online) (a) and (b) -- The gate voltage dependences of the
electron density and mobility, respectively, obtained at $V_g<1.7$~V within
framework of standard TTCC model. (c) -- The partial
electron and hole conductivities plotted against the gate voltage. The
parameters of electrons for $V_g>1.8$~V for all the panels are obtained as follows:
$n=1/e|R_H(0.03\text{~T})|$, $\mu_e=|R_H(0.03\text{~T})|\sigma$.}\label{f10}
\end{figure}

Thus, we assume that TTCC regime arises from  bands overlapping due to
non-monotonic  ``Mexican hat'' energy spectrum of the upper valence
subband and thus the data can be analyzed in the framework of  model of
laterally homogeneous 2D gas. It is naturally in this case to use the
classical hand-book formulae for two types of carriers to fit the
magnetic field dependences of $\rho_{xx}$ and $R_H$ (see, e.g.,
Ref.~\onlinecite{Blatt}). Such the fitting procedure has been performed
at low magnetic field, $B<0.3$~T,  with the use of densities and
mobilities of electrons and holes as the fitting parameters under
assumption that they are independent of magnetic field. The results of
the best for $V_g=1.1$~V and $V_g=-2$~V are shown in Fig.~\ref{f9}. It
is seen that this simple model quite well describes both dependences,
$\rho_{xx}(B)$ and $R_H(B)$. The gate voltage dependence of electron
density found by this manner at $V_g<1.7$~V is plotted in
Fig.~\ref{f10}(a) [points at $V_g>1.8$~V are obtained as
$1/e|R_H(0.03\text{~T})|$]. It is necessary to stress  that electron
contribution to the conductivity occurs down to the very large negative
gate voltages, $V_g\simeq -6$~V,  when the hole density becomes about
$5\times 10^{11}$~cm$^{-2}$ (see Fig.~\ref{f4}).

Let us consider the $V_g$ dependence of  density in greater detail. It
is rather complicated as evident from Fig.~\ref{f10}(a). There are
three $V_g$ intervals distinguished by the slopes. (i) At $V_g>1.8$~V,
when the Fermi level lies in the conduction band, the electron density
decreases with decreasing $V_g$ with the rate $\Delta n/\Delta V_g$ of
about $5.5\times 10^{10}$~cm$^{-2}$~V$^{-1}$. (ii) At $V_g=(1.0\ldots
1.7)$~V, the rate  is about ten times less, than that at $V_g>1.8$~V,
$\Delta n/\Delta V_g \simeq 6.0\times 10^{9}$~cm$^{-2}$~V$^{-1}$ [see
inset in Fig.~\ref{f10}(a)]. Quite apparently this feature results from
the fact that the holes appear at $V_g\simeq 1.8$~V, and the rate of
decrease of the Fermi energy is already determined by the hole density
of states, which is ten times larger than the electron one:
$dE_F/dV_g|_{V_g>1.8\text{~V}}=(1+m_h/m_e)\,
dE_F/dV_g|_{V_g<1.8\text{~V}}$. Thus, extrapolating the
$n$~versus~$V_g$ plot to $n=0$ one obtains that the electrons of the
conduction band have to disappear at $V_g\approx0$~V [see the solid
line in Fig.~\ref{f10}(a) and the inset in it]. (iii) At $V_g=(-6\ldots
+1)$~V, the electron contribution to the conductivity remains
essential, as it follows from analysis of the magnetic field
dependences of $R_H$ and $\rho_{xx}$, despite the fact that the
electrons in the conduction band are expected to disappear at
$V_g\approx0$~V. The data points in this range fall on the straight
line with the slope $\Delta n/\Delta V_g\simeq 1.5\times
10^{9}$~cm$^{-2}$~V$^{-1}$.

We  focus now attention on the gate voltage dependence of the electron
mobility [Fig.~\ref{f10}(b)]. As clearly  seen it is nonmonotonic;
sharp minimum is evident near $V_g\simeq 1.8$~V. At $V_g> 1.8$~V, when
the conductivity is determined by the conduction band electrons and the
holes in the valence band are absent, the mobility decreases with the
decreasing gate voltage.  This is natural because the decrease of the
electron density and hence the electron energy leads to the increase of
scattering probability independently of that short- or long-range
scattering potential determines the mobility. The increase of electron
mobility with decreasing $V_g$ at $V_g< 1.8$~V, where the electron
density carries on decreasing,  seems strange at first sight. However
it can be explained by holes appeared at these gate voltages, which
screen the potential of scatterers effectively due to the large
effective mass. The authors of Ref.~\onlinecite{Olshanetsky12} who
observed analogous behavior of the electron mobility come to the same
conclusion.

As was discussed above, the contribution of the conduction band
electrons to the conductivity has to disappear at $V_g\approx 0$~V so
presence of electron contribution up to the large negative $V_g$ seems
surprising. Broadly speaking, shunting conductivity channels of
technological nature can make such a type of contribution. They can be
located either above the quantum well or below it. If such channels are
situated above the well, i.e., between the quantum well and the gate
electrode, they should reveal themselves in the capacitance
measurements. However, no peculiarities in the voltage-capacity
characteristics caused by the depletion of these channels were observed
in our experiments. Besides, the contribution of the channels to the
conductivity has to be enhanced with the increasing positive gate
voltage due to the increase of electron density in these channels.
However, as Fig.~\ref{f10}(c) shows the electron contribution to the
conductivity  has a maximum at $V_g\simeq 0$~V and decreases with the
increasing positive gate voltage. If the shunting channels are located
below the well, their contribution should disappear at the negative
voltage applied to the back gate electrode. Our measurements performed
on the back-gated samples show that the voltage applied to the back-
and top gate electrodes changes the magnetic field dependences of $R_H$
and $\rho_{xx}$ analogously. Thus no conducting channels exist above or
below the quantum well.

\begin{figure}
\includegraphics[width=0.8\linewidth,clip=true]{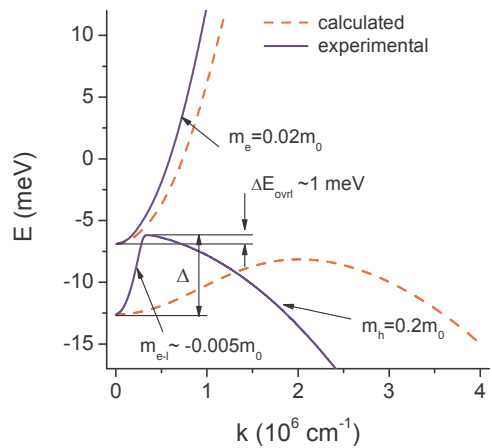}
\caption{(Color online) The dispersion $E(k)$ reconstructed from
the data analysis  as described in the text (solid lines) and calculated within
framework of isotropic six-band $kP$-model (dashed lines). }\label{f11}
\end{figure}

It remains to assume that the electron contribution at $V_g<0$~V is
caused by carriers in the well, namely, by the holes with electron-like
energy spectrum characterized by the negative effective mass in
vicinity of $k=0$. Thus, such data interpretation leads to the energy
spectrum sketched in Fig.~\ref{f11}. Qualitatively, it is close to the
calculated one, however quantitative difference is  significant. The
conduction and valence bands are overlapped. The value of overlapping
($\Delta E_\text{ovrl}$) can be estimated from the value of the
electron density at the gate voltage corresponding to appearance of the
hole contribution to the conductivity, $V_g\simeq 1.8$~V (see
Fig.~\ref{f10}). As seen from this figure $n\sim 1\times
10^{10}$~cm$^{-2}$ at $V_g\simeq 1.8$~V that gives $\Delta
E_\text{ovrl} \simeq n/\nu_e\sim 1$~meV if one uses the experimental
value of electron effective mass $m_e=0.02m_0$ to calculate  the
electron density of states $\nu_e$.

Experimentally, the hole energy spectrum is monotonic at $p\gtrsim
10^{11}$~cm$^{-2}$ that corresponds to $k=(2\pi p)^{1/2}\gtrsim
8\times10^5$~cm$^{-1}$. This conclusion follows immediately from the
two facts. First, the hole Hall density $p=1/eR_H$ is close to that
found from SdH oscillations. Second, the peaks of SdH oscillations are
shifted to the higher magnetic field when the gate voltage becomes more
negative (see Fig.~\ref{f5}). For $k<(7-8)\times 10^5$~cm$^{-1}$, the
hole energy spectrum is electron-like; there is narrow minimum in the
dispersion $E(k)$.  The depth of this minimum can be easily estimated
from the hole density wherein the electron contribution disappears:
$\Delta=p|_{n=0}/\nu_h$, where $p|_{n=0}$ is the hole density for the
gate voltage $V_g\simeq -6$~V [see Fig.~\ref{f10}(a)]. With the use of
experimental value of the hole effective mass, $m_h=0.2m_0$, and
$p|_{n=0}=p(V_g=-6\text{ V})\simeq (4 \ldots 5)\times
10^{11}$~cm$^{-2}$ (see Fig.~\ref{f4}), one obtains $\Delta\simeq
5$~meV. The effective mass of electron-like states of the valence band
is about $m_{e-l}\simeq -0.005\,m_0$. This estimate is obtained from
the depth of minimum $\Delta$ and from the amount of electron-like
states, $\simeq1\times 10^{10}$~cm$^{-2}$, estimated as $n$ in the
region $V_g=(0\ldots 1)$~V [see Fig.~\ref{f10}(a)]: $|m_{e-l}|=\pi
\hbar^2 n(V_g=1\text{ V})/\Delta$. Since   three types of carriers take
part in the transport at $V_g=(0\ldots 1)$~V, the $n$ values at these
gate voltages are obtained with considerably low accuracy. Therefore,
this estimate for $m_{e-l}$ should be considered as rough enough. In
Fig.~\ref{f11}, we reconstruct the energy spectrum using the parameters
$m_e$, $m_h$, $m_{e-l}$, $\Delta$ and $\Delta E_\text{ovrl}$ and
compare it with the spectrum calculated in framework of the $kP$-model.
One can see that $E$~versus~$k$ dependence in conduction band is close
to the theoretical one. The difference between the dispersion curves
$E(k)$ for the valence band is crucial.

We perceive that strong difference between the calculated and
reconstructed spectra indicates that discussion of alternative
approaches to the interpretation of the data is required. This will be
done below. Now let us examine the additional arguments in favor of the
energy spectrum depicted in Fig.~\ref{f11}. They can be obtained from
analysis of the data obtained in strong magnetic field.

As  noted above, the experimental dependences $\rho_{xy}(B)$ and
$\rho_{xx}(B)$ exhibit some peculiarities at the gate voltages near CNP
[in Figs.~\ref{f2}(a) and \ref{f2}(b), they are marked by the arrows].
The position of this minimum in the $(B,V_g)$ coordinates is shown in
Fig.~\ref{f5} by squares. Its origin directly relates to specific
features of the magnetic field quantization of the energy spectrum in
the quantum well with inverted spectrum.\cite{Schultz98} There are two
singular Landau levels responsible for that [see Fig.~1 in
Ref.~\onlinecite{Schultz98} and the inset in Fig.~\ref{f5}]. One of
them is the lowest level (h1, $n=0$) of the conduction band. Due to
complex nature of the band structure, its energy linearly decreases
with increasing magnetic field, while the other Landau levels of the
conduction band increase their energies as it takes place in usual
system. The second singular Landau level is the singular level (h2,
$n=2$) of the valence band, which starts at $B=0$ from the energy of h2
branch at $k=0$ and shifts upward with the increasing $B$. Unusual
behavior of these levels leads to a crossing of the conduction- and
valence band states at the critical value of the magnetic field,
$B=B_c$. The behavior of the crossing point with the change of the
quantum well width is investigated in Ref.~\onlinecite{Koenig07}. If
one extrapolates the calculation results\cite{Koenig07} to $d=20$~nm,
we obtain  $B_c\simeq (4-5)$~T, which is close to position of the
peculiarities evident in the $\rho_{xx}$- and $\rho_{xy}$~versus~$B$
dependences.

Thus, for the case when $n$ is slightly larger than $p$, the Fermi
level lies in the Landau level (h1, $n=0$) at $B\lesssim B_c$, while at
higher magnetic field, $B\gtrsim B_c$, it occurs in the Landau level
(h2, $n=2$) resulting in switching of the electron ground state and in
the peculiarity of $\rho_{xx}$ and $\rho_{xy}$. Existence of such
switching was discussed in recent paper.\cite{Raichev12}

The level (h2, $n=2$) reveals itself not only at low charge carrier
density. Really, inspection of Fig.~\ref{f5} shows that the data points
located above the dashed line fall on the straight lines, which are
well extrapolated to the common point $V_g\simeq 1.8$~V at $B=0$. Their
positions can be estimate from simple relation $n,p(V_g)=\pm ó BN/h$,
where $N$ is the number of occupied  Landau levels. Because
experimentally $n+p\simeq -5.5\times 10^{10}(V_g-1.8\text{
V})$,~cm$^{-2}$ (see Fig.~\ref{f4}), we obtain for $V_g^N$: $V_g^N=\pm
eBN/5.5\times 10^{10}h+1.8$~V. The solid lines in Fig.~\ref{f5} are
drawn according to this equation. It is seen that the separation
between the data points corresponding to $N=2$ and $3$ is well resolved
while the points with $N=4$ and $5$, $N=6$ and $7$, and so forth, are
merged. The points below the dashed line turn out to be shifted from
the lines. The reason is clear. When the level (h2, $n=2$) increasing
its energy with the growing magnetic field crosses the Fermi level at
magnetic field $B=B^\star$ (see inset in Fig.~\ref{f5}), the number of
Landau levels occupied by holes becomes larger on $1$ so that the
positions of minima below the dashed line in Fig.~\ref{f5} should be
described by $V_g^N=-eB(N+1)/5.5\times 10^{10}h+1.8$~V and, thus, the
oscillations turn out to be shifted (see solid lines in Fig.~\ref{f5}).
Thus, by and large this model describes the behavior of the oscillation
minima rather well. Some discrepancy is not surprising because this
simple model is valid when the overlapping between the Landau levels is
small, while the experimental data were obtained within the wide
magnetic field range involving both the Shubnikov-de Haas oscillations
and  the quantum Hall ranges.

The main conclusion, which follows from this consideration and from
Fig.~\ref{f5} is that the Landau level (h2, $n=2$)  at $B\to 0$ crosses
the Fermi level at  $V_g\simeq -6$~V, when the hole density is about
$4.5\times 10^{11}$~cm$^{-2}$. It means that the Fermi level lies below
the top of the valence band for a distance of $\sim 5$~meV. This value
is consistent with that estimated from the hole density at the gate
voltage corresponding to disappearance of the electron contribution to
the conductivity, that  supports the energy spectrum presented in
Fig.~\ref{f11} as well.

Let us now discuss alternative interpretations of data.  Analogous
heterostructures were investigated in
Refs.~\onlinecite{Kvon08,Olshanitsky09,Kozlov11,Kvon11}. Partially, our
results and results reported in these papers are overlapping. It
concerns in particular the value of the hole effective
mass,\cite{Kozlov11} existence of electron contribution to the
conductivity up to the hole density $(4-5)\times
10^{11}$~cm$^{-2}$.\cite{Kvon11} To interpret the results, the authors
of Refs.~\onlinecite{Kvon08,Olshanitsky09} suggest that there is an
overlap by about $5$~meV of the conduction band minimum at $k=0$ with
two symmetrical maxima of the valence band located at $k\neq 0$ in the
direction $[031]$ characterized by the effective masses close to
$0.2\,m_0$. This picture agrees well with the value of electron density
$n\simeq (4-5)\times 10^{10}$~cm$^{-2}$ (it is four-to-five times
higher than that in our case), at which the hole contribution to the
conductivity appears in samples investigated in
Ref.~\onlinecite{Olshanitsky09}. However this model of the energy
spectrum is in conflict with the fact that the hole density found from
the period of SdH oscillations coincides with that obtained from the
Hall effect. Because the valence band has two maxima in the spectrum,
the hole density obtained from the SdH experiments should be twice as
small as the Hall density. To resolve this contradiction the
authors\cite{Kvon11} assume that the spectrum is split by spin-orbit
interaction due to asymmetry of the quantum well. It is, however,
strange that this effect does not reveal itself in SdH oscillations.
The results\cite{Kvon08,Olshanitsky09} can be understood in framework
of our model, where the lines of constant energy are (nearly) circle
centered at $k=0$. However our results cannot be interpreted within
framework of the model suggested in
Refs.~\onlinecite{Kvon08,Olshanitsky09}. First, the hole contribution
to the conductivity in our samples appears at the electron density of
about $1\times 10^{10}$~cm$^{-2}$ instead of $5 \times
10^{10}$~cm$^{-2}$ in Ref.~\onlinecite{Olshanitsky09}. This corresponds
to the overlap value of about $1$~meV instead of $5$~meV. Second, the
spin-orbit splitting in our samples is evident in SdH effect at
$n\gtrsim 2\times 10^{11}$~cm$^{-2}$ and $p\gtrsim 4\times
10^{11}$~cm$^{-2}$. At lower densities its value is estimated as
$1$~meV or less, i.e., the valence band can be considered as unsplit
one under our experimental conditions.

Thus, our results and their interpretation lead us to the conclusions
on the energy spectrum of the HgTe quantum well which are inconsistent
with that obtained in framework of traditional $kP$-model. A most
surprising result is existence of narrow electron-like pit in the
center of the valence band with the depth of about $5$~meV
characterized by the very low effective mass $|m_{e-l}|\simeq
0.005\,m_0$.

\section{Conclusion}
\label{sec:concl}

We have studied the transport phenomena in HgTe single quantum well
with inverted energy spectrum. Consistent analysis of the magnetic
field dependences of the magnetoresistivity, the Hall coefficient, and
the SdH effect in the gated samples carried out over  the wide range of
the electron and hole densities including the charge neutrality point
leads us to the conclusion that the structure of the top of the valence
band is drastically different from that predicted in framework of
standard $kP$-approach. We obtain that the hole effective mass is equal
to approximately $0.2\,m_0$ at $k\gtrsim 0.7\times 10^{12}$~cm$^{-2}$
and practically independent of the hole density, while the theory
predicts negative (electron-like) effective mass up to $k^2\simeq
6\times 10^{12}$~cm$^{-2}$.  The experimentally obtained effective mass
near $k=0$, where the spectrum is electron-like, is close to
$-0.005\,m_0$, whereas the theory predicts the value less than $-0.1\,
m_0$. All this indicates that the further experimental and theoretical
investigations are needed to find the answer to the question of whether
the standard $kP$ model adequately describes the energy spectrum of
wide HgTe based single quantum well.

\section*{Acknowledgments}

This work has been supported in part by the RFBR (Grant Nos.
10-02-91336, 10-02-00481, and 12-02-00098).


%

\end{document}